\documentclass{article}
\usepackage{amssymb}
\begin{document}

{\bf ARE BLACK HOLE STARSHIPS POSSIBLE?}

\bigskip

By Louis Crane and Shawn Westmoreland, Kansas State University

\bigskip

\
\\
{\bf ABSTRACT:} {\it We investigate whether it is physically possible
  to build  starships or power plants using the Hawking radiation of an artificial
  black hole as  a power source. The proposal seems to be at the edge of
  possibility, but quantum gravity effects could change the picture. }

\bigskip
\
\\
{\bf I. Introduction}

\bigskip
According to Hawking, black holes (BHs) are not really black, but radiate what is approximately a black body thermal spectrum. The energy emitted is negligible unless
the BH is very small, in which case it becomes extremely energetic.

The purpose of this paper is to investigate whether it is possible to
build artificial BHs of the appropriate size, and to employ them in 
powerplants and starships. The conclusion we reach is that it is just
on the edge of possibility to do so, but that quantum gravity effects,
as yet unknown, could change the picture either way. 

We discuss designs for a family of machines which would
in combination realize our program. The machines are far beyond
current technological capabilities. The BH generator would be a gamma
ray laser with a lasing mass of order $10^{9}$ tonnes.

Nevertheless, we think the possibility should be studied
 carefully, because it would have profound consequences
for the distant human future, which no other proposal based on currently
known Physics could duplicate. 

This is only a preliminary study. Many questions which arise in this
program lead to calculations in general relativity which have not been
done. Whatever the other merits of our
proposal, we are confident it will pose many interesting problems for
classical and quantum relativity.

\bigskip

\
\\
{\bf II. The difficulty of interstellar flight}

\bigskip

{\bf A. Shielding}

\bigskip

The dreams of manned spaceflight of the fifties and sixties have
largely gone unrealized. This is to an important degree because it has been
discovered that cosmic radiation has much more serious medical 
consequences than was originally believed \cite{SciAm}.

 Substantial human presence
in space has only occurred in low Earth orbit near the equator, where
the earth's magnetic field shields us from the cosmic radiation.

Our visits to the moon were brief, and it is considered that a
prolonged human presence there would have to be underground,
necessitating the transport of massive ``earth'' moving equipment
\cite{SciAm}.

It is now known that 
any prolonged human presence deeper in space would need to be behind a shield
of the effective strength of two feet of lead, which would weigh 400
tonnes for a small capsule \cite{SciAm}.

It therefore becomes more economical to
think of a larger vessel, weighing many thousands of tonnes, in which
a group of people could live indefinitely. This possibility has not 
been very widely explored, particularly not in a practical direction,
because of the enormous energies involved
in accelerating such a body. 
\bigskip

{\bf B. Specific impulse} 

\bigskip

The distances between the stars are so great that practical travel between them
would require us to reach speeds comparable to the speed of
light. This is extremely difficult to do because very few processes
known to us release energies comparable to the masses of the matter
involved. Nuclear reactions, for example, release only a fraction of a
percent of the rest masses of the nucleii, so that an interstellar
vehicle powered by fission or fusion would have to carry many 
thousands of times the mass of its payload in fuel.

Coupled with the
shielding problem discussed in Section A above, this means an
interstellar voyage using nuclear energy would deplete the Earth's
resources of fissile or fusile materials to an intolerable degree.

Other than black hole radiation, which we study below, the only
process we know of which is sufficiently energetic is matter-antimatter annihilation. This has been proposed, but there are two
severe obstacles. 

The first is that the efficiency of antimatter production
in current accelerators is well below $10^{-7}$ (very few collisions produce a
trappable antiparticle) \cite{Forward}. Thus, making enough antimatter to propel a
starship would use up ten million times as much energy as our
proposal. The most optimistic projections of antimatter enthusiasts do
not produce an efficiency above $10^{-4}$, so that at best our
proposal is still ten thousand times more efficient.

The second obstacle is containment. A microscopic particle of
ordinary matter which drifted into the antimatter would cause an
explosion, scattering the antimatter into contact with the ship, and
destroying everything for millions of miles around. Any
electromagnetic force which held the antimatter in would also drive
normal matter in. One hears the suggestion that this could be solved by
``magnetic bottles,'' but magnetism is a force which acts
perpendicularly to the motion of a charged particle, and therefore
does not in any simple way form a bottle. Experiments in the  magnetic
confinement of plasma for millisecond intervals have been very frustrating.

Paramagnetic forces can repel matter, but they are extremely weak, and
treat matter and antimatter identically. So they would force normal
matter in as much as antimatter.

For these reasons an antimatter starship seems out of reach given
Physics as we know it. We could imagine surprises in future Physics
which would change this picture, but they seem remote. Dark
matter, for example, interacts neither with normal matter nor
antimatter.

Let us briefly contrast the idea of using antimatter with the
synthetic black hole we will discuss below. It is currently possible
to produce antimatter in extremely small quantities, while a synthetic
black hole would necessarily be very massive. On the other hand, the
process of generating a BH from collapse is naturally efficient, so it
would require millions of times less energy than a comparable amount
of antimatter or at least tens of thousands of times given some
optimistic future antimatter generator. 

As to confinement, a BH confines itself. We would need to avoid
colliding with it or losing it, but it won't explode. Matter striking a
BH would fall into it and add to its mass. So making a BH is extremely
difficult, but it would not be as dangerous or hard to handle as a
massive quantity of antimatter. 

Although the process of generating a
BH is extremely massive, it does not require any new Physics. Also, if a
BH, once created,  absorbs new matter, it  will radiate it, thus acting
as a new energy source; while antimatter can only act as a storage
mechanism for energy which has been collected elsewhere and converted
at extremely low efficiency.

(None of the
other ideas suggested for interstellar flight seem viable either.
The proposal for an interstellar ramjet turns out to produce more drag
than thrust, while the idea of propelling a ship with a laser beam
runs into the problem that the beam spreads too fast.)

Thus at this point, we are not even sure of a viable method for
sending very small probes to other stars. Sending human beings is much
harder, but in some sense it is what we really want, being what we are.

\bigskip

\
\\
{\bf III. Black Holes and Hawking radiation}

\bigskip

In the present paper, quantitative discussions on BHs  
will assume BHs of  the Schwarzschild type. 
This is the simplest type of BH; having mass but no 
angular momentum and no electric charge.  

\bigskip

{\bf A. Effective radius}
\bigskip

A Schwarzschild BH of mass $M$ has an ``effective radius" $R$ given by (e.g., \cite{Frolov}, pp. 15 - 16):
\begin{eqnarray}\label{radius}
R=\frac{2GM}{c^2}, 
\end{eqnarray} where $G$ is the gravitational constant 
and $c$ is the speed of light. Even though the geometry of spacetime is not Euclidean, the area of the event horizon 
of a  Schwarzschild BH (in a quiescent steady state) is given 
by the Euclidean formula $4\pi R^2$ (cf. \cite{Frolov}, p. 80). This justifies calling $R$ an  
``effective radius." The  effective radius provides a  
useful heuristic measure for  the ``size" of a BH. Coincidentally, 
Equation (\ref{radius}) can be deduced on Newtonian grounds 
whereby one seeks the radius $R$ of a   sphere of  mass $M$ such that 
the escape velocity at the surface of the sphere is equal to the speed of light.

\bigskip

{\bf B.  Hawking temperature}
\bigskip

In classical general relativity, BHs are regions of spacetime which are so peculiarly warped that matter and energy can flow in, but nothing can come out. However, quantum particles are known to  tunnel  through classically impenetrable  barriers. Indeed, Stephen Hawking has argued that quantum particles trapped inside of a BH can tunnel their way out to freedom. Thereby BHs emit a form of  thermal radiation called ``Hawking radiation."  Hawking calculated that, as observed from infinity, the effective temperature of a BH is proportional to the strength of the gravitational field  at its event horizon (cf. \cite{Frolov} p. 350). In the case of a Schwarzschild BH, with an effective radius of $R$, the temperature $T$ is given by (cf. \cite{Frolov} p. 351):
\begin{eqnarray}\label{HawkTemp}
T = \frac{\hbar c}{4\pi  k R},
\end{eqnarray} where $\hbar$ is the reduced Planck constant and $k$ is the Boltzmann constant. 

\bigskip

{\bf C. Power}
\bigskip

The luminosity, or total radiated power $P$, of a BH emitting Hawking radiation is somewhat subtle to compute. For a ``poor man's estimate," take the area of the BH to be $4\pi R^2$,  and take $T$ as given by Equation (\ref{HawkTemp}). The Stefan-Boltzmann black body law gives $P\propto 1/R^2$, where the constant of proportionality is about $7.86\times10^{-22} $ W$\cdot$m$^2$. This 
na\"ive approximation fails to take into account several important features however, such as  the   various different particle species that can be  created. It turns out  that  the total amount of radiated power  is, even at low energies, a bit underestimated in this way.  As the black hole gets smaller and more energetic, the error is compounded. Since we are interested in very energetic BHs, the estimate provided by the Stefan-Boltzmann law is a bit too 
na\"ive for   our present purposes.

Thereby we will use a more sophisticated calculation on the power of Hawking radiance which is based on work from  References \cite{Jane I} (MacGibbon and Webber) and \cite{Jane II} (MacGibbon).  MacGibbon calculated that the total amount of radiated  power $P$  produced by a BH with an effective radius of $R$  is given by (cf. \cite{Jane II} or \cite{Semiz}):
\begin{eqnarray}\label{power}
P=\frac{af(T)}{R^2},
\end{eqnarray} where $a$ is a constant with a value of about $1.06\times10^{-20}$ W$\cdot$m$^2$  and $f(T)$ is a temperature-dependent numerical factor having to do with the various particle species emitted.  One always has $f(T)\geq1$. 
A precise computation of  $f(T)$  requires detailed knowledge about particle physics.   

In the Standard Model, one has $f(T)\lesssim 15.4$ (cf. \cite{Jane II}). 
MacGibbon \cite{Jane II} argues that $f(T)\lesssim 100$ in supersymmetric and technicolor models.

Since the constant $a=1.06\times 10^{-20}$ W$\cdot$m$^2$ has  such  a tiny magnitude, it follows from Equation (\ref{power}) that in order for  a BH to yield a significant amount of power via Hawking radiation, the BH must be of ``subatomic" dimensions; having an effective radius much less than $10^{-10}$ m. 
Such ``subatomic black holes" (SBHs), as we shall call them,
 are not known to occur naturally in the present-day
 universe. Hypothetical  ``primordial black holes," which
   would now be in their final (explosive) stages of evaporation, have not been
   detected.
   
   Not all of the power generated  by a SBH may be practically   useful. Hot BHs  create massive particles, so some of the emitted ``energy" actually goes into  the ``rest mass" of these particles.  Also, a  certain percentage  of the radiated energy  will end up in the form of  neutrinos, particles which  interact only very weakly with matter.   However,  for the very hot BHs which are of interest to the present paper, this ``neutrino drain" is not terribly significant and most of the radiated power is emitted in more readily accessible  channels \cite{Semiz}.

Following MacGibbon \cite{Jane II}, we will approximate $f(T)$ by the formula:
\begin{eqnarray}\label{MacGibbonf(T)}
f(T) & =& 1.569 + 0.963\exp\left(\frac{-0.10\textrm{ GeV}}{kT}\right)
+0.569\left[\exp\left(\frac{-0.0234\textrm{ GeV}}{k T}\right)
\right.\nonumber\\
&&\left.+ 6\exp\left(\frac{-0.066\textrm{ GeV}}{kT}\right) +3\exp\left(\frac{-0.11\textrm{ GeV}}{kT}\right)\right.\nonumber\\
&&\left. + \exp\left(\frac{-0.394 \textrm{ GeV}}{kT}\right) + 3\exp\left(\frac{-0.413 \textrm{ GeV}}{kT}\right) \right.\nonumber\\
&&\left.+ 3\exp\left(\frac{-1.17\textrm{ GeV}}{kT}\right) + 3\exp\left(\frac{-38 \textrm{ GeV}}{kT}\right)\right],
\end{eqnarray}
which should hold reasonably well in the case of a BH with an energy-equivalent temperature $kT$ that is above $0.06$ GeV but less than 100 GeV \cite{Jane II}. The effective radius of such a BH is between about  0.16 and  260 attometers. ($1$ attometer $=10^{-18}$ meters.)

We have taken the liberty to make a slight correction to the last term in Equation (\ref{MacGibbonf(T)}). This term  corresponds to the contribution of the top quark. The correction is not significant, but at the time of MacGibbon's paper \cite{Jane II}, the mass of the top quark was unknown and MacGibbon assumed its  mass to be about 100 GeV. Today we know that the top quark has a mass of about 171 GeV \cite{PDG} and thereby we have adjusted Equation (\ref{MacGibbonf(T)}) accordingly. 

It is known that Equation (\ref{MacGibbonf(T)}) slightly underestimates the quark contribution, treating them as particles with $\pm1$ charge rather than fractional charge, but the correction is not very significant \cite{Jane II}. Also, Equation (\ref{MacGibbonf(T)}) does not include  the $W^\pm$ and $Z^0$ emissions (which are only worth noting at temperatures of about 15 GeV or greater \cite{Jane II}), but again this does not have a very significant impact on $f(T)$ (see \cite{Jane II} or \cite{Semiz}).

More generally, one can approximate the function $f(T)$ as  a sum $f(T) = \sum_{i} f_i(T)$, where $i$ ranges over all of the particle species  that can be directly created by a BH of temperature $T$. The terms $f_{i}(T)$ can be estimated via:
\begin{eqnarray}\label{spectra}
f_i(T) = h_{i}n_{i}\exp\left(-bq_i^2-\frac{m_ic^2}{\beta_i kT}\right).
\end{eqnarray} This  equation for $f_i(T)$  is taken from Appendix B of  Semiz's paper \cite{Semiz}, in which some of the results from Reference \cite{Jane II} are nicely summarized. The values of  $h_i$ and $\beta_i$ depend on spin and are listed in Table \ref{table1}. The constant $b$ has a value of about $0.034$. The variable  $q_i$ is the charge (in electron units) of the $i$th particle species, $n_i$ is the number of internal degrees of freedom for the particle, and $m_i$ is the mass of the particle.

\begin{table}[h]
\begin{center}
\begin{tabular}{c|c|c|c|c}
spin$_i$& 0 &1/2 & 1 &  2\\
\hline
$h_i$&0.267 & 0.147 & 0.060  & 0.007\\
\hline
$\beta_i$&2.66&4.53&6.04&9.56\\
\end{tabular}
\caption{ The values of $h_i$ and $\beta_i$ according to spin (as given in  References \cite{Jane II} and  \cite{Semiz}).}\label{table1}
\end{center}
\end{table}

\bigskip

{\bf D. Life expectancy}
\bigskip

By mass-energy conservation, a BH which emits an infinitesimal amount $dE$ of energy must decrease its mass by $dM = -dE/c^2$ (cf. \cite{Jane II}). Thereby, the rate at which mass ``leaks out" of a BH is:
\begin{eqnarray}\label{massdecrease}
-\frac{dM}{dt}=\frac{1}{c^2}\cdot\frac{dE}{dt}= \frac{P}{c^2}.
\end{eqnarray}
By substituting Equations (\ref{radius}) and (\ref{power}) into Equation (\ref{massdecrease}) we obtain:
\begin{eqnarray}\label{radiusdecrease}
\frac{dR}{dt}=-\frac{2Gaf(T)}{c^4R^2}.
\end{eqnarray} Hence, a BH with an initial effective radius of $R_0$ has a life expectancy of  $L$ where:
\begin{eqnarray}\label{lifetime0}
L=\int_0^{R_0}\frac{c^4 R^2}{2Gaf(T)}dR.
\end{eqnarray} By feeding on mass at a rate comparable to the mass-leakage rate $P/c^2$, a BH can last well beyond its life expectancy.

As an isolated BH shrinks smaller and grows hotter, the value of $f(T)$ does not in general remain constant. Generously taking $f(T)\leq 100$, the integral in  Equation (\ref{lifetime0}) can be estimated as:
\begin{eqnarray}\label{range1}
\frac{c^4 R_0^3}{600Ga}\leq L\leq \frac{c^4 R_0^3}{6Gaf(T_0)},
\end{eqnarray} where $f(T_0)$ denotes the minimum value of $f(T)$ that an isolated BH of radius $R_0$ will radiate through over   its remaining lifetime. 

A SBH of radius $0.6$ attometers has an energy-equivalent temperature of about $26.2$ GeV. Using Equation (\ref{MacGibbonf(T)}), we find that at this temperature, $f(T)\approx 12.5$. If  $f(T)$ is a  nondecreasing function of $T$ for the temperatures that an isolated SBH with a    radius of  $0.6$ attometers will achieve over its remaining lifetime, then by (\ref{range1}) one finds that the life expectancy of such a BH is less than about 1.04  years. Indeed,  MacGibbon \cite{Jane II} reports that a BH with a mass of $4\times10^{11}$ grams  (such a BH has a radius of about $0.6$ attometers) has  a life expectancy of approximately 1 year. 

Let $R_Y =0.6$ attometers and let $Y=1$  year. Given  that a BH with a radius of $R_Y$ has a life expectancy of about 1 year, we get that for a BH of  radius $R_0$:
\begin{eqnarray}\label{l-y}
L-Y\approx \int_{R_Y}^{R_0} \frac{c^4 R^2}{2Gaf(T)}dR.
\end{eqnarray}  If the radius  $R_0$  is between 0.16 and 260 attometers, then Equation (\ref{MacGibbonf(T)}) applies and  approximates $f(T)$ by an increasing function of $T$.  Therefore,  the integral in (\ref{l-y}) can be estimated as:
\begin{eqnarray}\label{lifetime}
\frac{c^4 }{75Ga}(R_0^3 - R_Y^3)\lesssim L- Y 
\lesssim \frac{c^4}{6Gaf(T_0)} (R_0^3-R_Y^3),
\end{eqnarray} where $T_0$ is the  temperature of a BH with a radius of $R_0$.

A more sophisticated approximation for $L$ can be carried out by making use of the ``mass-squared average" of $f(T)^{-1}$ over the lifetime of an isolated BH (see MacGibbon \cite{Jane II} for details). However,  the estimate for $L$ as given by the range (\ref{lifetime})  will suffice for the purposes of the present paper. 

\bigskip

{\bf E. A note on charged and spinning black holes}
\bigskip

For details on how things change if the SBH is electrically charged  or spinning, we recommend that  the reader consult an authoritative reference such as the book by Frolov and Novikov  \cite{Frolov}. Note that if an isolated SBH is initially endowed with an electric charge, then it will quickly, and almost completely, radiate this charge away  (see \cite{Frolov}, p. 398). Spinning SBHs  radiate more powerfully than non-spinning ones of equal mass (see \cite{Frolov}, pp. 399 - 400, or Page \cite{PageIII}). However,  the angular momentum of an isolated SBH is  rapidly dissipated (see \cite{Frolov}, p. 399, or  \cite{PageIII}).

\bigskip

\
\\
{\bf IV. Theoretical Feasibility}

\bigskip
In this section we want to discuss whether the Physics of black holes
discussed above, together with the laws of Physics of matter as we
know them, make it possible to produce artificial BHs which would be
useful, either as power plants or as starships.

Since the mass of a black hole decreases with its radius, while its
energy output increases and its life expectancy decreases, this is a
delicate question.

\
\\
{\bf List of criteria:}
{\it We need a black hole which}
\bigskip
\
\\
{\it 1. has a long enough lifespan to be useful,}
\bigskip
\
\\
{\it 2. is powerful enough to accelerate itself up to a reasonable fraction of the speed of light in a reasonable amount of time,}
\bigskip
\
\\
{\it 3. is small enough that we can access the energy to make it,}

\
\\
{\it 4. is large enough that we can focus the energy to make it,}
\bigskip
\
\\
{\it 5. has mass comparable to a starship.}

\bigskip

We could easily imagine that this would be impossible. Somewhat
surprisingly, it turns out that there is a range of BH radii, which
according to the semiclassical approximation, fit these
criteria.

Using the formulae from the section above, we find that a black hole with a radius of a few 
 attometers at least roughly meets the list of
criteria (see Appendix). Such BHs would have mass of the order of 1,000,000 tonnes,
and  lifetimes ranging from decades to centuries. A  high-efficiency square solar panel a few hundred km on each side, in a circular orbit about the sun at a distance of 1,000,000 km,  
would  absorb enough energy in a year to produce one such BH. 

A BH with a life span on the order of a century would emit 
enough energy to accelerate itself to relativistic velocity in a
period of decades. If we could let it get smaller and hotter before feeding 
matter into it, we could get a better performance.

In Section V below, we discuss the plausibility of creating SBHs with a very large spherically converging gamma ray laser.  A radius of 1 attometer corresponds to the wavelength of a  gamma
ray with an energy of about   1.24 TeV. Since the wavelength of the Hawking
radiation is $8\pi^2$ times the radius of the BH, the Hawking temperature of a BH with this radius  is on the order of 16 GeV, 
within the limit of what we could hope
to achieve technologically.

Now the idea that the wavelength of the radiation should match the
radius of the BH created is very likely pessimistic. The collapsing
sphere of radiation would gain energy from its self-gravitation as it
converged, and there is likely to be a gravitational self-focussing.

This is a problem that can be studied using standard techniques from
classical general relativity in which Einstein's equation is coupled
to Maxwell's equations in vacuum, or ``electrovac.'' We intend to
investigate this in the future.

Thus it seems that making an artificial black hole and using it to
drive a starship is just possible, because the family of BH solutions
has a ``sweet spot.'' This seems not to have been remarked before in
the literature, except for the earlier work of one of us.

Perhaps this is just a cosmic coincidence. In the epilogue, we discuss
possible philosophical ramifications of this observation.

\bigskip

\
\\
{\bf V. Four Machines}

\bigskip

Now we discuss how a technology for implementing our proposal could be
implemented. These devices are far beyond current technology, but we
think they are possibly
capable of being implemented ultimately if a future industrial society
were determined to do so.

\bigskip

{\bf A. The black hole generator}
\bigskip

In a previous paper by the first author \cite{Crane}, it was proposed that a SBH 
could be artificially  created by firing a huge number of  gamma rays from a spherically converging laser. The idea is to pack so much energy into such a small space that a  BH will form.  An advantage of using photons is that, since they are bosons,  there is no Pauli exclusion principle to worry about. Although a laser-powered black hole generator presents huge engineering challenges, the  concept appears to be physically sound according to classical general relativity. The Vaidya-Papapetrou metric shows  that an imploding spherically symmetric shell of ``null dust" can form a black hole  (see, e.g., \cite{Frolov}, p. 187, or Joshi \cite{Joshi} for further details).

Since photons  have null stress energy just like null dust,
 a black hole should form if a large aggregate of photons interacts classically with the gravitational
field. As long as we are discussing regions of spacetime that are many
orders of  magnitude larger than the Planck length, we should be
outside of the regime of quantum gravity and classical theory should
be appropriate. However, the assumption of spherical symmetry is
rather special, and an investigation into the sensitivity of the
process to imperfections in symmetry is an interesting problem for
classical general relativity. If a high degree of spherical symmetry
is required, then this could pose serious engineering challenges.

Since a nuclear laser can convert on the order of $10^{-3}$ of its
rest mass to radiation, we would need a lasing mass of order $10^9$
tonnes to produce the pulse. This should correspond to a mass of order
$10^{10}$ tonnes for the whole structure (the size of a small asteroid). Such a structure would be
assembled in space near the sun by an army of robots and built out of
space-based materials. It is not larger than some structures human
beings have already built. The precision required to focus the
collapsing electromagnetic wave would be of an order already possible
using interferometric methods, but on a truly massive scale.

This is clearly extremely ambitious, but we do not see it as impossible.

\newpage

{\bf B. The drive} 

\bigskip
Now we would like to discuss how to use a SBH to drive a starship. We
need to accomplish 3 things.

\bigskip

{\bf Design requirements for a BH starship}

\bigskip

{\it 1. use the Hawking radiation to drive the vessel}

\bigskip

{\it 2. drive the BH at the same acceleration}

\bigskip

{\it 3. feed the BH to maintain its temperature}

\bigskip

Item 3 is not absolutely necessary. We could manufacture a SBH, use it
to drive a ship one way, and release the remnant at the
destination. However this would limit us greatly as to performance,
and be very disappointing in the powerplant application discussed below.

We shall discuss these three problems in outline only here; at the
level of engineering they will each require an extended discussion.

It is not hard to see how we might satisfy requirement 1. We simply
position the SBH at the focus of a parabolic reflector attached to the
body of the ship. Since the SBH will radiate  gamma rays and a mix of
particles and antiparticles, this is not simple. The proposal has been
made in the context of antimatter rockets, to make a gamma ray reflector
out of an electron gas \cite{Sanger}.

It is not clear if this is feasible (e.g., \cite{Forward}).

Alternatively, we could allow the gamma rays to escape and direct only the charged particle part of the Hawking radiation  (cf. \cite{Forward}), although this produces a less capable ship. To improve the performance, we could  add a thick layer of matter which would absorb the gamma rays, reradiate in optical frequencies, and focus the resulting light rays. An absorber which stops only  gamma rays heading towards the front of the ship and allows the rest to escape out the back  causes gamma rays to radiate from the ship asymmetrically. In this way, even the escaping non-absorbed gamma rays contribute some thrust (cf. \cite{Smith} or \cite{WIP}). 
Modulo safety concerns, one would not want the  absorber to be too massive. An extremely massive absorber could burden the mass of vehicle so much that the extra thrust it helps to deliver  does not lead to an improved acceleration. 

%Alternatively, we could add a thick layer of
%matter which would absorb the gamma rays, reradiate in optical
%frequencies, and focus the resulting light rays. It is also possible
%to allow the gamma rays to escape and direct only the charged particle
%part of the Hawking radiation, although this produces a less capable
%ship.   

Yet another idea for the utilization of gamma ray energy is to exploit pair production phenomena.  By interacting with the electric field of atomic nuclei,  high energy gamma rays can be converted into charged  particle-antiparticle pairs such as electrons and positrons.  These particles  can  be directed by electromagnetic fields. It is not  likely that even half of the gamma ray energy can be utilized in this manner  however (see Vulpetti \cite{Vulpetti83}, \cite{Vulpetti85}).

It might be advantageous  to use the Hawking radiation  to energize a secondary working substance which can then be ejected as exhaust   (as is done in  thermal  and ion rockets). However, the working substance must be ejected at relativistic speeds so that  the specific impulse will be high enough  for interstellar travel.

The most optimistic approach is to solve requirements 2 and 3 together
by attaching particle beams to the body of the ship behind the BH and
beaming in matter. This would both accelerate the SBH, since BHs ``move when
you push them''(see  \cite{Frolov} p270), 
and add mass to the SBH, extending the lifetime.

The delicate thing here is the absorption cross section for a particle
going into a BH. We intend to investigate this question in the future. If simply
aiming the beam at the SBH doesn't work, we can try forming an
accretion disk near the SBH and rely on particles to tunnel into
it. Alternatively, we could use a small cluster of SBHs instead of just
one to create a larger effective target, charge the SBH etc. It is
also possible that because of quantum effects SBHs have larger than
classical radii, due to the analog of zero point energy. 

This point must remain as a challenge for the future.

\bigskip

{\bf C. The powerplant}
\bigskip

This has already been proposed by Hawking (see \cite{BHOT} pp. 108 - 109). We simply surround the SBH with a spherical shield, and use it to
drive heat engines. (Or possibly use gamma ray solar cells, if such
things be.) This would have an enormous advantage over solar electric
power in that the energy would be dense and hence cheaper to accumulate.

The 3 machines here really form a tool set. Without the drive, getting
the powerplant near Earth where we need it would be very
difficult. Without the generator, it would require the good fortune to
find a primordial SBH to implement the proposal.

\bigskip

{\bf D. The self-driven generator}

\bigskip 
The industry formed by our first 3 machines would not yet be really
mature. To fully tap the possibilities we would need a fourth machine,
a generator coupled to a family of SBHs which could be used to charge
its laser. Assuming we can feed a SBH as discussed above, we would
then have a perpetual source of SBHs, which could run indefinitely on
water or dust or whatever matter was most convenient.

A civilization  equipped with our four machine tool set would be almost
unimaginably energy rich. It could settle the galaxy at will.

\
\\
{\bf  VI. Open questions; quantum corrections}

\bigskip

The reader has no doubt by now observed that a great many questions in
this proposal are left open. We can mention the self-focussing   of a
focussing electromagnetic wave, and the possible effects of gravitational
lensing and magnification on the various aspects of our problem. 

Another open issue has to do 
 with calculating  the amount of gravitational radiation 
 that would be produced by an SBH in a starship or other piece of SBH-powered technology (see below).

When we come to quantum gravity effects, the questions are almost
endless. The effect of a Planck scale cutoff on Hawking radiation,
tunnelling, and modifications of our formulas by quantum self-energy
are some of the obvious ones.

We think the proposal will be interesting for researchers in quantum
gravity as a source of problems for their theories with a practical
flavor, which could actually be studied in the distant future. The shortage of
experiments in quantum gravity is so dire that even gedanken
experiments would be helpful.

\bigskip
\
\\
{\bf VII. A New Approach to SETI}

\bigskip

As Freeman Dyson \cite{Dyson} has pointed out, when working with speculative ideas about technologies that appear to be physically possible, but beyond our grasp, we are faced with two options. We 
can either speculate on what humanity might achieve in the distant future or we can
speculate on what a hypothetical extraterrestial civilization might have
already done. The first line of thought postpones the possibility of any tangible results  indefinitely into the  future. On the
other hand, the second line of thought raises a legitimate scientific
question that could be settled by contemporary or near-term astronomical observations. We can ask ourselves what an extraterrestrial  BH starship or other high technology would look like from  Earth and  proceed to look for it, if possible.

A SBH capable of driving a starship produces Hawking radiation which  ultimately gives rise to    gamma rays, neutrinos, antineutrinos, electrons, positrons,  protons, and antiprotons  \cite{Jane II}. Gamma ray telescopes are already in use and thereby one might think that a careful search through the gamma ray sky  could conceivably  turn up evidence of an extraterrestrial  starship (cf. \cite{Harris}).  However, gamma rays produced by  a SBH in a distant starship might be extremely difficult to detect if the starship is very energy-efficient and has well-collimated exhuast jets. 

A BH starship using the technology we are proposing would emit
gravitational radiation at nuclear frequencies. Current gravitational
radiation detection experiments are optimized for much lower
frequencies, and would not detect it.

We propose building gravity wave detection devices of a different
design, with sensitivity in wavelengths of nuclear order. Since the
wavelengths to be detected are much shorter, and since the energy
outputs would be so high and the potential sources could be much
closer, it would be possible to make detectors with a much shorter arm
length, build them on Earth, and use them to scan for a signal. This
would be a high risk high gain experiment; a positive signal would be
an indication of extraterrestrial civilization.

It is also possible that such a detector would discover something
truly unexpected; it is a new way to probe the universe.

We intend to undertake calculations as to the feasibility of such
detectors in the future.

Meanwhile, here is a very rough back-of-the-envelope calculation of possible relevance. If  spin 2 massless gravitons $g$ exist, then, due to the Hawking effect, an isolated non-rotating electrically neutral BH will radiate them
at a total power of $P_{g} = af_{g}(T)/R^2$. Since the graviton has two degrees of freedom ($n_g=2$) (see, e.g., \cite{Feynman} pp. xi,  39), we get from Equation 
(\ref{spectra}) and Table \ref{table1},  that $f_{g}(T) = 0.014$. 
An isolated SBH with an effective radius between $1$ and $6$ attometers 
will therefore radiate gravitons at a power on the order of 0.004 to 0.1 petawatts. This can be thought of as a lower estimate on the rate at which 
a SBH will radiate gravitational energy. If the SBH is being accelerated 
or perturbed in some other way by an external agent,  then one would expect that  more gravitational 
radiation will be emitted from the SBH. On the other hand, the gravitational radiation could not
exceed the rate at which energy was being fed into the SBH, which in a
steady state  stardrive would equal the Hawking radiation from all
particle species, which is on the order of 3 to 130 petawatts for SBHs with radii in our proposed range. 

We note that the graviton emission of a rapidly spinning SBH   could be a few orders of magnitude   more powerful than the non-spinning case considered above (see  \cite{PageIII}).   

\bigskip

\
\\
{\bf VIII. Conclusions}

\bigskip

Quantum gravitational corrections could make this proposal easier or
impossible. Within our current understanding of Physics, this
proposal could make greater concentrations of energy available for human use
than anything else we currently know. It may even allow us to
go to the stars in person, rather than simply sending miniaturized probes.

The
proposal we are making should be pursued as far as possible, and in as
optimistic a spirit as the facts permit, because it allows a completely
different and vastly wider destiny for the human race. We should not
underestimate the ingenuity of the engineers of the future.

As an aside, let us note that building a single SBH would be the
ultimate particle Physics experiment. As the remnant BH radiated it
would go through every temperature up to the Planck scale. Careful
study of its radiation would tell us exactly what particle types exist
in nature, beyond the energy of any imaginable accelerator. Perhaps a
future society would carry out such an experiment, even if all 
proposals for applications fail.

\bigskip

\
\\
{\bf IX. Epilogue: The meduso anthropic principle}

\bigskip

The origin of this proposal is very peculiar. The first author was
reviewing the work of Lee Smolin, which was later published in a book 
entitled \emph{The Life of the Cosmos} \cite{Smolin}. 
Professor Smolin proposed that the universe we see was
only one of many universes, and that new universes arise from old ones
whenever a black hole is produced. This then leads to an evolutionary
process for universes in which universes with an unusually high number
of stars were selected for.
 
The first author proposed that the evolutionary process of universes
should include life. This is possible if successful industrial
civilizations eventually produce black holes, and therefore baby universes.
                               
Note that if successful industrial civilizations only trapped already
existing BHs, it would not alter the number of baby universes a
universe produced, so that no evolutionary loop would result.

This led us to consider the possibility of producing artificial BHs,
and to explore how they might be useful.

The meduso-anthropic principle (as this proposal was named) is at
least falsifiable, in the sense that if it turned out artificial BHs
were completely  impossible or useless then the evolutionary cycle
universes-civilizations-black holes -baby universes could never happen.

The result of our feasibility calculation is much too tenuous to be
considered a proof of the meduso-anthropic principle. It is not in
fact clear at all that black holes create baby universes, as the
maximal analytic continuations of the standard BH solutions to
Einstein's equation suggest. 

Nevertheless, it 
is a bit eerie that only through this line of thought did we
consider the possibility of synthetic BH creation. We are the first
and almost the
only authors to our knowledge to consider this. The only other author we
are aware of is Semiz \cite{Semiz}, who wrote: ``...we would have to
either find small black holes, possibly primordial, or manufacture
them by means as yet unknown,'' in a rather popular discussion on BH
powerplants and stardrives, which appeared after our first paper.

\bigskip

\
\\
{\bf Acknowledgements}

\bigskip

The authors would like to thank Adam Crowl for his constructive
comments. The first author was supported on FQXi grant BG0522, and
minigrant BG1006.

\bigskip

\
\\
{\bf Appendix: Finding the ``sweet spot"}

\bigskip

In this Appendix, we show that a BH with a critical radius of a few attometers  at least roughly meets the criteria listed in Section IV. 

Using the semiclassical formulae from Section III, we have  calculated the Hawking temperatures, total power outputs, mass-leakage rates, and life expectancies for various SBHs of radii between 0.16 and 10 attometers. The results are presented in Table \ref{table2}. The figures reported are the rounded results of complete calculations. In each case,  the value of $f(T)$, as approximated by Equation (\ref{MacGibbonf(T)}), is used to estimate the power outputs, mass-leakage rates, and life expectancies. Since it is known that Equation (\ref{MacGibbonf(T)}) slightly underestimates $f(T)$, it follows that  the power outputs and mass-leakage rates are slightly underestimated and the life expectancies are slightly overestimated.

\begin{table}[h]
\begin{center}
\begin{tabular}{lllllll}
$R$ (am) & $M$ (Mt) & $kT$ (GeV)& $f(T)$& $P$ (PW)& $P/c^2$ (g/sec)  & $L$ (yrs) \\
\hline
0.16 & 0.108 &98.1&13.3 & 5519& 61400&$\lesssim0.04$ \\
0.3 & 0.202 &52.3&13.0 & 1527& 17000&$\lesssim0.12$ \\
0.6 & 0.404 & 26.2&12.5 & 367 & 4090 & $1$\\
0.9 & 0.606 & 17.4&12.2 &160 & 1780 & $3.5$\\
1.0 & 0.673 &15.7& 12.1&129 & 1430 & $5$\\
1.5 & 1.01 &10.5&11.9 & 56.2 & 626 & $16-17$\\
2.0 & 1.35 & 7.85&11.8 &31.3 & 348 & $39-41$\\
2.5& 1.68 &6.28 & 11.7& 19.8 & 221 & $75-80$\\
2.6 &  1.75& 6.04&  11.7&  18.3&204 &$85-91$\\
2.7 & 1.82& 5.82&11.7 & 16.9 & 189 & $95-102$\\
2.8 &1.89 & 5.61&  11.6 & 15.7 & 175& $106-114$\\
2.9 & 1.95&5.41 & 11.6&14.6  & 163 & $118-127$\\
3.0 & 2.02 & 5.23&11.6 & 13.7 & 152 & $130-140$\\
&&& \\
5.8 & 3.91& 2.71 &11.1  & 3.50 & 38.9  & $941-1060$\\
5.9 & 3.97&2.66 &11.1 & 3.37 & 37.5 & $991-1117$\\
6.0 & 4.04& 2.62& 11.1&3.26  &36.2  & $1042-1177$\\
&&&\\
6.9 & 4.65& 2.28 &10.9  &2.43 & 27.1 & $1585-1814 $\\
7.0 & 4.71 &2.24& 10.9& 2.36 & 26.2 & $1655-1897$\\
&&&\\
10.0 &6.73 & 1.57&10.5 & 1.11& 12.3 & $4824-5763$ \\
\hline
\end{tabular}
\caption{This table shows effective radii $R$ (in attometers), masses $M$  (in millions of tonnes),  Hawking temperatures $kT$ (in GeVs), estimated $f(T)$-values, estimated total power outputs $P$ (in petawatts), estimated mass-leakage rates $P/c^2$ (in grams per second), and estimated life expectancies  $L$ (in years). Note that the power-to-mass ratio $P/M$  (not explicitly shown)  is very high - especially for smaller SBHs.}
\label{table2} 
\end{center}
\end{table}

According to Table \ref{table2}, a BH with a  radius of 1 attometer or less has a mass-leakage rate exceeding 1 kilogram per second. Such a BH will radiate very powerfully, but have a short life expectancy. Unless the BH is  fed mass-energy at a rate comparable to the mass-leakage rate, it will expire quickly. If it is not possible to feed such a BH then the BH must be safely disposed of in a timely manner. According to the semiclassical approximation, a BH  ``explodes" at the end of its lifetime. Such an explosion is powerful by terrestrial standards, but not by astronomical standards. A BH that explodes at a distance of 1 AU from Earth poses no danger. Simply on the basis that $f(T)\geq1$, Inequality (\ref{range1}) shows  that when the BH reaches a radius of about $8\times 10^{-22}$ m, its life expectancy is less than a second. Generously assuming that  $f(T)\leq  100$, Equation (\ref{power}) tells us that such a BH has a  total radiated power of less than about $1.66\times 10^{24}$ W. The sun itself  radiates  about 230 times more powerfully than this.  According to the semiclassical approximation, the BH will go on to radiate more and more powerfully until it finally expires, but Earth will  only be exposed to this more intense radiation for an insignificant fraction of a second.  Moreover, if the ``explosion" occurs behind the sun, then the sun will act as a shield and Earth will be exposed to even less Hawking radiation.  The upshot is that SBHs can be safely disposed of when necessary.

\bigskip
{\bf A. What BHs are long-lived enough and powerful enough for interstellar travel?}

\bigskip

SBHs of radii less than 1 attometer are incredibly powerful. Note however that the life expectancy of a BH with a radius of 0.16 attometers is less than about 2 weeks. In order for such a BH to last  significantly longer than that, an external agent must force-feed it mass-energy at a rate of many kilograms per second.  
As we have emphasized throughout the text,  It is unknown whether
SBHs,  since they are so very small, can feed on anything at all,  let
alone many kilograms  per second.  If  SBHs with radii on the order of $0.1$ 
attometers could be force-fed  at sufficiently high rates, by using a
feeding system whose mass is small compared to the mass of the SBH,
then it is not hard to believe that  such SBHs could be very  useful as power sources for  starship propulsion systems - if their power can be harnessed efficiently.  In the following
however, we will assume that SBHs cannot be fed. Even in this ``worst
case scenario," it turns out that SBHs could  still turn out to be
useful for interstellar travel. 

We note that guiding a BH is less difficult than feeding it, because
it is only necessary to scatter radiation off the BH to impart momentum
to it. If even this is impossible, it is hard to see how to build any
drive at all.

About the fastest type of interstellar voyage that  human beings  could physically tolerate  would be a one-way trip from Earth to Alpha Centauri (a distance of just over 4 light years \cite{Shepherd}) which accelerates at a proper acceleration of 1 g for the first half of the voyage and  decelerates  at  1 g for the second half. In this way, the travelers arrive at  Alpha Centauri with zero relative speed. The trip would only take about 3.5 years from the perspective of the travelers (thanks to Special Relativity).

From Table \ref{table2}, a BH with a life expectancy of about 3.5 years has a radius of about 0.9 attometers. Unless SBH lifetimes can be significantly extended via feeding, a manned interstellar vehicle powered by an on-board SBH requires SBHs of at least this initial size (and most likely quite larger).

Conceivably, unfed SBHs of radii less than 0.9 attometers, having less than 3.5 year life expectancies, could be used to rapidly accelerate  interstellar robotic probes to relativistic speeds.  Robotic probes  do not necessarily need to ``stop" and could tolerate much larger accelerations than humans. The problem of navigating such objects could be difficult however. 

The  SBH would have to be ejected (or otherwise escaped from) before  it explodes.

A SBH with a radius of 0.9 attometers has a mass of about 606,000 tonnes and a power output of about 160 petawatts. Over a period of only  20 days a 160 petawatt power source emits  enough energy to accelerate  606,000 tonnes up to about $10\%$ the speed of light. Of course, it is unrealistic to suppose that the  emitted energy can be converted into kinetic energy with $100\%$ efficiency, but even if the conversion occurs with an efficiency of only $10\%$, it only takes 10 times longer to deliver the requisite  kinetic energy. 

If we cannot use SBHs that are more powerful than 160 petawatts (i.e., smaller than 0.9 attometers) for manned starships, then it may not be  feasible to use SBHs as power sources for   interstellar rockets that maintain \emph{both}  constant proper accelerations of 1 g \emph{and}  extremely  high exhaust speeds (near $c$).   An ideal rocket undergoing a proper acceleration of 1 g, whose exhaust consists of perfectly collimated  light-like radiation (exiting the vehicle at $c$), needs to consume about  3,000 petawatts per million tonnes of vehicle mass  \cite{Shepherd}. It turns out that the ratio of power dissipation to  acceleration becomes more tolerable when lower exhaust speeds are used \cite{Shepherd}, but interstellar travel demands very high exhaust speeds.

 On the other hand, interstellar travel   does not necessarily require large accelerations. Small accelerations sustained over long periods of time suffice.

SBHs with radii of a few attometers  are more long-lived and could be used for long-term  voyages. Since a ``larger SBH has less power," and is therefore less capable of acceleration, one would accelerate the ship somewhat weakly - at least during the initial part of the journey.  As the journey progresses,  the (unfed) SBH shrinks in size and  becomes more powerful. 

Assuming that SBHs cannot be fed, an interstellar voyage of  100 years  would need a SBH with a life expectancy of at least 100 years. According to Table \ref{table2}, a SBH with a radius of about 2.7 attometers has a life expectancy of about 100 years (but remember that the life expectancies in the table are somewhat overestimated, as noted above).  A  mission which needs SBH power for 1,000 years would require a SBH with a radius of about 5.9 attometers. Now  a 1,000 year voyage is not very appealing, and we take this to be a generous upper limit on the amount of time that a single interstellar mission might take. (A SBH with a life expectancy of over 1,000 years would probably not be suitable in a stardrive anyhow - see below.) On the other hand, if  the destination is  reached well before the BH evaporates, then the BH could  serve temporarily as a power plant after it has served in a stardrive.

According to Table \ref{table2}, a 100 year BH has a mass of about 1,820,000 tonnes, and radiates at about 17 petawatts. It takes about 1.5 years for a 17 petawatt power source to emit enough energy  to accelerate 1,820,000 tonnes to $10\%$ the speed of light.  If the emitted energy can only be converted into kinetic energy at an efficiency of  $10\%$, then it takes about 15 years. If the efficiency is only $5\%$, then it takes approximately 30 years, but even this is  good enough for long-term interstellar travel. (We are ignoring, for the moment, the fact that  the BH radiates more powerfully over time.)

  A 1,000 year BH  has a radius of about $5.9$ attometers, a mass of about 3,970,000 tonnes, and radiates at about 3 petawatts. It takes such an object almost 20 years to emit the amount of energy needed to  accelerate  3,970,000 tonnes to 10\% the speed of light. Assuming that the emitted energy could be converted into kinetic energy with only $10\%$ efficiency, it would  take closer to  200  years.

The above calculation  hints that a SBH with a life expectancy of over 1,000 years (and a  radius in excess of about 6 attometers) would not serve adequately as the power source for a starship.  Even the adequacy of a  1,000 year BH is somewhat doubtful.    

Indeed, it appears that a  SBH with a radius of 10 attometers would be highly inadequate. Such a BH has a life expectancy of a few millennia, a  mass of about 6,730,000 tonnes, and radiates about 1 petawatt. It takes about a century for a 1 petawatt power source to emit enough energy to accelerate 6,730,000 tonnes up to 10\% the speed of light (and this even ignores the fact that the emitted energy could not be converted into kinetic energy with perfect efficiency). It seems to be a safe bet that one would not want to use such a BH in a stardrive.

The upshot is that SBHs with effective radii   between roughly 1 and  6 attometers could be adequate power sources for starship propulsion systems - if their power can be  harnessed  efficiently. The smaller the BH, the more powerful it is. 
However, one must either be able to arrive at one's  destination before the BH evaporates completely or one must be able to feed the BH  in such a way as  to prolong its life. In the most optimistic scenario, where one can feed the SBH very efficiently, one could have very capable manned starships  driven by extremely powerful   SBHs with  radii smaller than our suggested range of 1 to 6 attometers.

 \bigskip
 
{\bf B. Are these BHs so \emph{small} that we can \emph{access} the energy to make them?} 
 
 \bigskip
 
We now consider the question of whether a SBH  can plausibly be made using  energy sources that could at least eventually be at our disposal. The most abundant energy source in the solar system is the sun, which has a luminosity of about 
 $3.8427\times 10^{26}$ W \cite{PDG}. Following a transition from a global Earth-based economy to an interplanetary Solar System-based economy (whereby off-Earth energy resources are utilized), a reasonable fraction of the sun's total power will be at our disposal. The sun releases the equivalent of 2 million tonnes of energy in  less than half a second, which is enough energy to make a BH with an effective radius of a few attometers. 
 
A perfectly efficient square solar panel with each side measuring about 370 km, in a circular orbit about the sun at a distance of 1,000,000 km, would absorb enough energy in one year to create a  BH with an effective radius of about $2.2$ attometers.  As discussed in the text, one possible mode of production for an initial set of SBHs is to use a  very massive nuclear laser. Such a laser can be constructed near the sun so that it can be ``charged up" by solar energy before getting switched on.

 \bigskip
 
{\bf C. Are these BHs  so \emph{big} that we can \emph{focus} the energy to make them?} 
 
 \bigskip

This is the ``dual" side of the problem discussed in part B above.  Suppose that 
we are to synthesize a BH with an effective radius between $1$ and $6$ attometers  by focusing a spherically converging gamma ray
laser at a single point. If  the photons coming from the laser need to have wavelengths of no more than about the critical radius  of the BH, then  we need to use gamma ray photons  of  energies between about  210 and 1240 GeV. As discussed in the text, it could turn out that wavelengths significantly larger than the critical radius would be alright. If we can use gamma rays having energies roughly matching the  Hawking temperature  the SBH to be synthesized, then we need gamma rays having energies  of the order 
3 - 16 GeV. These are comparable to wavelengths within the Compton
radii of   
nucleii, hence could be technically possible.
 
 \bigskip
 
{\bf D. Do these  BHs have masses comparable to that of a starship?} 
 
 \bigskip

Since we are still pretty far away from building starships, the ``mass of a typical starship" is  a matter of speculation.  It is  a safe bet however that a starship would have to be very massive. It must shield its passengers and sensitive technological equipment from hazardous cosmic radiation. It would also have to accommodate a large population for a potentially long time (from decades to perhaps a century or more). 

SBHs with radii ranging from 1 to 6  attometers have masses ranging from 673,000 to 4,040,000 tonnes. We conjecture that  a starship  powered by a \emph{single} SBH would transport payloads having a total mass of somewhat \emph{less} than this. Much larger payloads could be transported by very large starships powered by several SBHs.

\end{document}